\RequirePackage{lineno}

\documentclass[twocolumn,aps,prl,amsmath,amssymb,superscriptaddress,showpacs]{revtex4}

\usepackage{graphicx, color}
\usepackage{xspace}
\usepackage{lineno}
\usepackage[labelsep=period]{caption}

\newcommand{\cbar}{\ensuremath{\bar{c}}\xspace}

\newcommand{\tbar}{\ensuremath{\bar{t}\xspace}}
\newcommand{\bbar}{\ensuremath{\bar{b}\xspace}}
\newcommand{\pbar}{\ensuremath{\bar{p}\xspace}}

\newcommand{\ccbar}{\ensuremath{{c\cbar}}\xspace}
\newcommand{\bbbar}{\ensuremath{{b\bbar}}\xspace}
\newcommand{\ttbar}{\ensuremath{{t\tbar}}\xspace}

\newcommand{\ppbar}{\ensuremath{{p\pbar}}\xspace}

\newcommand{\de}{\ensuremath{\mathrm{d}}}

\newcommand{\ra}{\ensuremath{\rightarrow}}

\newcommand{\pt}{\ensuremath{p_{T}}\xspace}

\newcommand{\et}{\ensuremath{E_{T}}\xspace}

\newcommand{\met}{\ensuremath{\et\!\!\!\!\!\!\!/\,\,}\,\,\xspace}

\newcommand{\gev}{\ensuremath{\mathrm{GeV}}\xspace}

\newcommand{\gevcsq}{\ensuremath{\mathrm{GeV}/c^{2}}\xspace}
\newcommand{\tev}{\ensuremath{\mathrm{TeV}}\xspace}
\newcommand{\invfb}{\ensuremath{\mathrm{fb}^{-1}}\xspace}

\newcommand{\ie}{i.e.,\ }

\newcommand{\Zj}{$Z$ + jets\xspace}

\newcommand{\Wj}{$W$ + jets\xspace}
\newcommand{\Wlf}{$W+$LF\xspace}
\newcommand{\Whf}{$W+$HF\xspace}

\newcommand{\btagging}{$b$-tagging\xspace}

\newcommand{\pyth}{{\sc pythia}\xspace}

\newcommand{\alp}{{\sc alpgen}\xspace}
\newcommand{\Lint}{\ensuremath{\int\!\mathcal{L}\de t}\xspace}

\newcommand{\MC}{Monte Carlo\xspace}

\newcommand{\ntag}{\ensuremath{\mathrm{n}_{tag}}\xspace}
\newcommand{\njet}{\ensuremath{\mathrm{n}_{jet}}\xspace}

\newcommand{\topxsecassym}{\ensuremath{7.64^{+0.57}_{-0.54}}\xspace} 
\newcommand{\topxsecfull}{\ensuremath{7.64 \pm 0.57~\mathrm{(stat + syst)} \pm 0.45~\mathrm{(luminosity)}}\xspace}
\newcommand{\wbbkassym}{\ensuremath{1.39^{+0.28}_{-0.22}}\xspace}

\newcommand{\wcckassym}{\ensuremath{0.83^{+0.90}_{-0.71}}\xspace} 
\newcommand{\wckassym}{\ensuremath{1.68^{+0.34}_{-0.32}}\xspace} 
\newcommand{\wqqkassym}{\ensuremath{0.98^{+0.34}_{-0.25}}\xspace}
\newcommand{\ewkassym}{\ensuremath{1.10^{+0.10}_{-0.10}}\xspace}

\newcommand{\qcdkassym}{\ensuremath{0.82^{+0.26}_{-0.26}}\xspace}

\begin{document}

\title{Measurement of the Top Pair Production Cross Section in the Lepton + Jets 
Channel Using a Jet Flavor Discriminant}

\affiliation{Institute of Physics, Academia Sinica, Taipei, Taiwan 11529, Republic of China} 
\affiliation{Argonne National Laboratory, Argonne, Illinois 60439, USA} 
\affiliation{University of Athens, 157 71 Athens, Greece} 
\affiliation{Institut de Fisica d'Altes Energies, ICREA, Universitat Autonoma de Barcelona, E-08193, Bellaterra (Barcelona), Spain} 
\affiliation{Baylor University, Waco, Texas 76798, USA} 
\affiliation{Istituto Nazionale di Fisica Nucleare Bologna, $^{aa}$University of Bologna, I-40127 Bologna, Italy} 
\affiliation{University of California, Davis, Davis, California 95616, USA} 
\affiliation{University of California, Los Angeles, Los Angeles, California 90024, USA} 
\affiliation{Instituto de Fisica de Cantabria, CSIC-University of Cantabria, 39005 Santander, Spain} 
\affiliation{Carnegie Mellon University, Pittsburgh, Pennsylvania 15213, USA} 
\affiliation{Enrico Fermi Institute, University of Chicago, Chicago, Illinois 60637, USA}
\affiliation{Comenius University, 842 48 Bratislava, Slovakia; Institute of Experimental Physics, 040 01 Kosice, Slovakia} 
\affiliation{Joint Institute for Nuclear Research, RU-141980 Dubna, Russia} 
\affiliation{Duke University, Durham, North Carolina 27708, USA} 
\affiliation{Fermi National Accelerator Laboratory, Batavia, Illinois 60510, USA} 
\affiliation{University of Florida, Gainesville, Florida 32611, USA} 
\affiliation{Laboratori Nazionali di Frascati, Istituto Nazionale di Fisica Nucleare, I-00044 Frascati, Italy} 
\affiliation{University of Geneva, CH-1211 Geneva 4, Switzerland} 
\affiliation{Glasgow University, Glasgow G12 8QQ, United Kingdom} 
\affiliation{Harvard University, Cambridge, Massachusetts 02138, USA} 
\affiliation{Division of High Energy Physics, Department of Physics, University of Helsinki and Helsinki Institute of Physics, FIN-00014, Helsinki, Finland} 
\affiliation{University of Illinois, Urbana, Illinois 61801, USA} 
\affiliation{The Johns Hopkins University, Baltimore, Maryland 21218, USA} 
\affiliation{Institut f\"{u}r Experimentelle Kernphysik, Karlsruhe Institute of Technology, D-76131 Karlsruhe, Germany} 
\affiliation{Center for High Energy Physics: Kyungpook National University, Daegu 702-701, Korea; Seoul National University, Seoul 151-742, Korea; Sungkyunkwan University, Suwon 440-746, Korea; Korea Institute of Science and Technology Information, Daejeon 305-806, Korea; Chonnam National University, Gwangju 500-757, Korea; Chonbuk National University, Jeonju 561-756, Korea} 
\affiliation{Ernest Orlando Lawrence Berkeley National Laboratory, Berkeley, California 94720, USA} 
\affiliation{University of Liverpool, Liverpool L69 7ZE, United Kingdom} 
\affiliation{University College London, London WC1E 6BT, United Kingdom} 
\affiliation{Centro de Investigaciones Energeticas Medioambientales y Tecnologicas, E-28040 Madrid, Spain} 
\affiliation{Massachusetts Institute of Technology, Cambridge, Massachusetts 02139, USA} 
\affiliation{Institute of Particle Physics: McGill University, Montr\'{e}al, Qu\'{e}bec, Canada H3A~2T8; Simon Fraser University, Burnaby, British Columbia, Canada V5A~1S6; University of Toronto, Toronto, Ontario, Canada M5S~1A7; and TRIUMF, Vancouver, British Columbia, Canada V6T~2A3} 
\affiliation{University of Michigan, Ann Arbor, Michigan 48109, USA} 
\affiliation{Michigan State University, East Lansing, Michigan 48824, USA}
\affiliation{Institution for Theoretical and Experimental Physics, ITEP, Moscow 117259, Russia}
\affiliation{University of New Mexico, Albuquerque, New Mexico 87131, USA} 
\affiliation{Northwestern University, Evanston, Illinois 60208, USA} 
\affiliation{The Ohio State University, Columbus, Ohio 43210, USA} 
\affiliation{Okayama University, Okayama 700-8530, Japan} 
\affiliation{Osaka City University, Osaka 588, Japan} 
\affiliation{University of Oxford, Oxford OX1 3RH, United Kingdom} 
\affiliation{Istituto Nazionale di Fisica Nucleare, Sezione di Padova-Trento, $^{bb}$University of Padova, I-35131 Padova, Italy} 
\affiliation{LPNHE, Universite Pierre et Marie Curie/IN2P3-CNRS, UMR7585, Paris, F-75252 France} 
\affiliation{University of Pennsylvania, Philadelphia, Pennsylvania 19104, USA}
\affiliation{Istituto Nazionale di Fisica Nucleare Pisa, $^{cc}$University of Pisa, $^{dd}$University of Siena and $^{ee}$Scuola Normale Superiore, I-56127 Pisa, Italy} 
\affiliation{University of Pittsburgh, Pittsburgh, Pennsylvania 15260, USA} 
\affiliation{Purdue University, West Lafayette, Indiana 47907, USA} 
\affiliation{University of Rochester, Rochester, New York 14627, USA} 
\affiliation{The Rockefeller University, New York, New York 10065, USA} 
\affiliation{Istituto Nazionale di Fisica Nucleare, Sezione di Roma 1, $^{ff}$Sapienza Universit\`{a} di Roma, I-00185 Roma, Italy} 

\affiliation{Rutgers University, Piscataway, New Jersey 08855, USA} 
\affiliation{Texas A\&M University, College Station, Texas 77843, USA} 
\affiliation{Istituto Nazionale di Fisica Nucleare Trieste/Udine, I-34100 Trieste, $^{gg}$University of Trieste/Udine, I-33100 Udine, Italy} 
\affiliation{University of Tsukuba, Tsukuba, Ibaraki 305, Japan} 
\affiliation{Tufts University, Medford, Massachusetts 02155, USA} 
\affiliation{University of Virginia, Charlottesville, VA  22906, USA}
\affiliation{Waseda University, Tokyo 169, Japan} 
\affiliation{Wayne State University, Detroit, Michigan 48201, USA} 
\affiliation{University of Wisconsin, Madison, Wisconsin 53706, USA} 
\affiliation{Yale University, New Haven, Connecticut 06520, USA} 
\author{T.~Aaltonen}
\affiliation{Division of High Energy Physics, Department of Physics, University of Helsinki and Helsinki Institute of Physics, FIN-00014, Helsinki, Finland}
\author{B.~\'{A}lvarez~Gonz\'{a}lez$^w$}
\affiliation{Instituto de Fisica de Cantabria, CSIC-University of Cantabria, 39005 Santander, Spain}
\author{S.~Amerio}
\affiliation{Istituto Nazionale di Fisica Nucleare, Sezione di Padova-Trento, $^{bb}$University of Padova, I-35131 Padova, Italy} 

\author{D.~Amidei}
\affiliation{University of Michigan, Ann Arbor, Michigan 48109, USA}
\author{A.~Anastassov}
\affiliation{Northwestern University, Evanston, Illinois 60208, USA}
\author{A.~Annovi}
\affiliation{Laboratori Nazionali di Frascati, Istituto Nazionale di Fisica Nucleare, I-00044 Frascati, Italy}
\author{J.~Antos}
\affiliation{Comenius University, 842 48 Bratislava, Slovakia; Institute of Experimental Physics, 040 01 Kosice, Slovakia}
\author{G.~Apollinari}
\affiliation{Fermi National Accelerator Laboratory, Batavia, Illinois 60510, USA}
\author{J.A.~Appel}
\affiliation{Fermi National Accelerator Laboratory, Batavia, Illinois 60510, USA}
\author{A.~Apresyan}
\affiliation{Purdue University, West Lafayette, Indiana 47907, USA}
% ...
\collaboration{CDF Collaboration\footnote{With visitors from $^a$University of Massachusetts Amherst, Amherst, Massachusetts 01003,
$^b$Istituto Nazionale di Fisica Nucleare, Sezione di Cagliari, 09042 Monserrato (Cagliari), Italy,
$^c$University of California Irvine, Irvine, CA  92697, 
$^d$University of California Santa Barbara, Santa Barbara, CA 93106
$^e$University of California Santa Cruz, Santa Cruz, CA  95064,
$^f$CERN,CH-1211 Geneva, Switzerland,
$^{ii}$CERN, PH-TH Geneva, Switzerland, 
$^g$Cornell University, Ithaca, NY  14853, 
$^h$University of Cyprus, Nicosia CY-1678, Cyprus, 
$^i$University College Dublin, Dublin 4, Ireland,
$^j$University of Fukui, Fukui City, Fukui Prefecture, Japan 910-0017,
$^k$Universidad Iberoamericana, Mexico D.F., Mexico,
$^l$Iowa State University, Ames, IA  50011,
$^m$University of Iowa, Iowa City, IA  52242,
$^n$Kinki University, Higashi-Osaka City, Japan 577-8502,
$^o$Kansas State University, Manhattan, KS 66506,
$^p$University of Manchester, Manchester M13 9PL, England,
$^q$Queen Mary, University of London, London, E1 4NS, England,
$^r$University of Melbourne, Victoria 3010, Australia,
$^s$Muons, Inc., Batavia, IL 60510,
$^t$Nagasaki Institute of Applied Science, Nagasaki, Japan, 
$^u$National Research Nuclear University, Moscow, Russia,
$^v$University of Notre Dame, Notre Dame, IN 46556,
$^w$Universidad de Oviedo, E-33007 Oviedo, Spain, 
$^x$Texas Tech University, Lubbock, TX  79609, 
$^y$Universidad Tecnica Federico Santa Maria, 110v Valparaiso, Chile,
$^z$Yarmouk University, Irbid 211-63, Jordan,
$^{hh}$On leave from J.~Stefan Institute, Ljubljana, Slovenia, 
}}
\noaffiliation

\begin{abstract}
  We present a new method to measure the top quark pair production cross section 
  and the background rates
  with data corresponding to an integrated luminosity of $2.7$~\invfb 
  from \ppbar collisions at $\sqrt{s} = $1.96$~\tev$ collected with the CDF~II Detector. 
  We select events with a single electron or muon candidate, 
  missing transverse energy, and at least
  one $b$-tagged jet. We perform a simultaneous fit to a jet flavor discriminant across nine
  samples defined by the number of jets and $b$-tags.
  An advantage of this approach is that
  many systematic uncertainties are measured {\it in situ} and inversely
  scale with integrated luminosity.
  We measure a top cross section of $\sigma_{\ttbar} = \topxsecfull$~pb.
\end{abstract}


\maketitle

Since its discovery in 1995~\cite{topdisccdf,topdiscdzero}, much has been learned
about the top quark through analyses of \ppbar collisions. 
Top quarks are produced in pairs through the
strong interaction and 
each top quark decays dominantly to a $W$ boson and
a $b$ quark, followed by the $W$ decaying either to a pair of quarks
(which form jets) or a lepton and a neutrino.
This paper describes a measurement of the top-antitop pair production cross
section, $\sigma_{t\bar{t}}$, in the $\ppbar \ra \ttbar \ra \ell \nu qq^\prime \bbbar$ 
channel at a center-of-mass energy, $\sqrt{s} =1.96~\tev$ using a new methodology to
constrain background contributions and systematic effects, resulting in an improved
sensitivity.

The jets which originate from the bottom quarks in the final state provide an opportunity
to select events which are more likely to have come from top quark decays than 
from other processes. A $b$-tagging algorithm takes advantage of the characteristics---largely 
the secondary vertex displaced from the primary vertex---that distinguish heavy flavor (HF) 
jets from charm and light flavor (LF) jets~\cite{btagalgo}.
This algorithm 
allows us to reduce the backgrounds from \Wj processes, which can mimic the top
decay signature, and was the basis for several previous measurements of the top cross 
section~\cite{topxseccdf, topxsecdzero}.

While requiring the event to have at least one $b$-tagged jet reduces the backgrounds, it does
not eliminate them. 
It is important to estimate the amount of $W$ boson production with
associated jets from heavy flavor, which is theoretically difficult and a source of systematic 
uncertainties for measurements of the cross section as well as the mass. 
Here we reduce this systematic uncertainty and constrain the \Whf background 
by performing a fit to the data which includes regions dominated by \Wj.

Since the $b$-tagging algorithm can incorrectly tag light flavor jets as $b$ jets, it is 
advantageous to apply an additional discriminant to $b$-tagged jets
to further separate processes with jets from bottom, charm, and light flavor. 
This flavor separator is a neural network whose output,
on a statistical basis, discriminates between $b$-quark, $c$-quark, and light-flavor jets. 
The flavor separator uses 25 variables to output 
a single number indicating how likely a jet is a $b$ jet, where the 
invariant mass of the secondary vertex has the most separation power. The flavor
separator was calibrated using data control samples~\cite{KIT}.

In this paper, we use a flavor separator for the first time in the measurement
of the \ttbar cross-section. In order to constrain the background contributions, we perform the fit 
to the flavor discriminant in nine samples defined by the number of jets, \njet (1, 2, 3, 4, 
or $\geq$ 5),
and number of $b$-tagged jets, \ntag (1 or $\geq$ 2).
Events with one or two jets are dominated by \Wj, whereas events with three or more jets are
largely \ttbar. Events with two $b$ tags are dominated by $W\bbbar$ and \ttbar, 
whereas events with a single $b$ tag are predominantly $W$+charm and \Wlf.
Previous methods selected events with three or more jets in order to reduce
the largest background from \Wj processes~\cite{topxseccdf, topxsecmethtwo, topxsecdzero}. 
This new method instead
constrains the background contribution of the \Wj processes in the region
with three or more jets by measuring the contributions in the regions with
one and two jets.


We use a data sample corresponding to $2.7$~\invfb of integrated luminosity, 
collected from February 2002 through April 2008 using the CDF~II
detector~\cite{cdf}, an approximately cylindrically symmetric detector located at the
Tevatron collider.  
CDF~II is a general-purpose device; 
the central drift chamber provides charged-particle tracking,
while the silicon system provides excellent vertex and impact
parameter resolution, both of which are important for identifying 
bottom quarks.  Electromagnetic and hadronic
calorimeters are located outside the tracking chambers, and provide
identification of electrons and jets. At the outermost
layer of the detector sit the muon drift chambers which provide muon
identification.

We select events with a $W$ candidate decaying leptonically to either an electron or muon.
We require at least one jet and exactly one lepton candidate both with transverse energy $\et >
20~\gev$ and 
pseudorapidity $|\eta| <~ 2.0$~\cite{coords}. 
We require that at least one jet is
$b$-tagged, and that there is at least $20~\gev$ of missing transverse
energy, \met, in the event. 
To reduce QCD backgrounds, we require the transverse mass of the $W$,
$m_T^W = \sqrt{2(p_T^\ell p_T^{\nu} - p_x^\ell p_x^{\nu} - p_y^\ell 
  p_y^{\nu})}$, 
  to be at least $10~\gevcsq$ for muons, and at least
$20~\gevcsq$ for electrons.  Electron samples have a larger QCD background contamination 
than muon samples, so there we also require the \met to satisfy
$\frac{\met}{\sqrt{\overrightarrow{E}_T^{\rm uncl} \cdot \hat{\met}}} 
> -0.05~m_T^W~\mathrm{(in \gevcsq)}+ 3.5$,
where the denominator is the square root of the amount of unclustered energy in the
direction of the missing transverse energy~\cite{qcdveto}. 


In addition to QCD multi-jet processes, the final state in this analysis can be mimicked by several
other processes. \Wj processes are by far the largest source of backgrounds. Single top 
production, di-boson production, and \Zj processes --- collectively referred to as 
electroweak (EW) processes --- also contribute. All but the QCD multi-jet backgrounds are modeled
with \MC simulations; the QCD backgrounds are estimated using a data-driven approach.
Events that pass the selection criteria, though with the lepton candidate 
failing any two identification cuts, are mostly QCD multi-jet processes, and this
sample is used to model the background from these QCD processes. 
The normalization of the QCD background is estimated from a fit to the 
missing \et distribution for each sub-sample, without the missing \et requirement.

\MC samples are employed to estimate acceptances for the signal and backgrounds, and 
to model relevant distributions used in the fits described below. All of the \MC
samples employed were generated using either 
\pyth v6.216~\cite{pythiamanual} (the \ttbar 
and diboson samples),  
{\sc madgraph}~\cite{madgraph} (the single top sample), or \alp v$2.10^\prime$~\cite{alpgen} with
generator-to-reconstructed-jet matching~\cite{mlmmatch1, mlmmatch2} 
and \pyth v6.326 for showering (the \Wj and \Zj samples). 
The \ttbar signal \MC sample was generated with the CTEQ5L~\cite{cteq} parton 
distribution functions (PDFs)
assuming a top mass of $m_t = 175~\gevcsq$. 
All samples are processed through a detailed simulation of the CDF~II detector response, 
after which they are treated in the same manner as the data events.
Each of the samples is divided based on the number
of jets and $b$ tags, and made into 
templates---binned distributions of the flavor separator output.


The measurement is 
accomplished as a fit of the flavor separator distribution performed
simultaneously in the nine data sub-samples.
Results are obtained by maximizing a binned Poisson likelihood which incorporates
templates from each of the \ttbar, \Wj, electroweak, and
QCD processes.
The templates are combined after 
initializing them to the predicted yield for data corresponding to an integrated luminosity of 
2.7~\invfb, where
the initialization factors are functions of 
cross sections, tagging efficiencies, and energy scales which are
all parameters in the fit. 
%
The overall normalization of each template is floated in the fit --- a 
single overall normalization factor is used for each process --- and 
the primary result of the fit is a set of those 
normalizations: the \ttbar cross section
and relative normalizations, $K_p$, to the standard model expectations 
for \Wj, electroweak, and QCD components.

\begin{figure}[htbp]
 \centering
    \resizebox{0.9\linewidth}{!}{\includegraphics{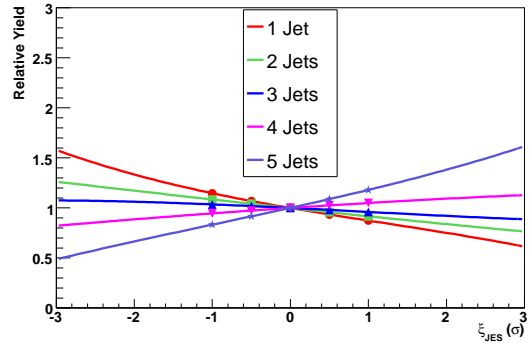}}
  \caption{(Color) The function which parameterizes the effect of the jet energy scale
  on the 1-tag templates of the \ttbar sample. Each jet and tag bin for each process 
  in each subsample has a different function for each source of systematic uncertainty. The
  x axis is in units of the systematic shift.}
  \label{fig:polynoid}
\end{figure}

Template normalizations also include
functions, $P_x(i,j,\xi_x)$, that parametrize the effect of a
source of systematic uncertainty, $x$, in the sub-sample with $\ntag = i$ and $\njet = j$, 
as a function of the relative shift, $\xi_x$, of quantity $x$, in units of
the uncertainty on $x$.
A separate function is employed for each process in each sub-sample for each source 
of systematic uncertainty; an example function is shown in Fig.~\ref{fig:polynoid}. 
This leads to a total of 12 parameters in the fit---seven normalizations
of the samples ($\sigma_{\ttbar}$, 
$K_{W\bbbar}$, $K_{W\ccbar}$, 
$K_{Wc}$, 
$K_{W+LF}$, 
$K_{EW}$, and $K_{QCD}$), and five systematic uncertainty parameters ($\xi_{Btag}$, 
$\xi_{Mistag}$, $\xi_{I/FSR}$, $\xi_{Q^2}$, $\xi_{JES}$).

Systematic uncertainties in this measurement can affect both the
normalizations and the shapes of the
templates.  The rate uncertainties are naturally included in the fit
via the $P_x(i,j,\xi_x)$ functions, and these systematic uncertainties are
reflected by the total fit error.  To account for each shape uncertainty, we
generate an additional set of templates with the variable in question
changed, 
re-run the fit, and take the difference in the result as the
uncertainty.

We vary the \btagging efficiency, mistag rate, and the jet 
energy calibration~\cite{jes} by their uncertainties for all simulated samples. 
%
%
Initial- and final-state radiation (ISR/FSR) are processes in which
gluons are radiated before or after the collision, respectively.  The
uncertainty arises due to ISR/FSR leading to a larger or smaller number
of jets in the event. To account for this, we make additional sets of \ttbar 
templates with more or less ISR and FSR as compared to the normal
settings; the different settings are constrained by studies of Drell-Yan 
production~\cite{dyisrfsr}. 
The systematic uncertainty associated to the choice of the 
renormalization and factorization scales is estimated by varying these 
scales between half and twice their default values, as indicated in \cite{alpgen}. This variation
also accounts for differences in ISR/FSR in \Wj processes.
%

The uncertainty due to the choice of the algorithm used to generate the parton 
shower was determined by comparing the results obtained using \pyth 
and {\sc herwig}~\cite{herwig}.
%
%
%
%
We account for uncertainties in our
modeling of the QCD template shape by using electronlike signals associated with  
multiple tracks, rather than electrons that fail identification cuts, to make templates.
The flavor separator has a correction factor applied to match its mistag rate to
the one observed in data; to account for this uncertainty, we examine
templates without this factor applied. The models describing color
reconnection---\ie the QCD cross-talk between the decay products of the top
quarks---are not known precisely, so we account
for this uncertainty by comparing two different models.
We take an 
uncertainty of 0.6\% on the top
cross section due to the PDFs, and 
an uncertainty of 0.5\% due to the beam position and lepton identification efficiency.
We take a conservative 2\% uncertainty due to the PDFs on the \Wj results. 
The measured luminosity has an uncertainty of 5.9\%. 

The total normalizations of the \ttbar and $W\bbbar$ components in the fit are given by 
\begin{align}
N_{\ttbar}^{pred}(i,j) &= \sigma_\ttbar \cdot L \cdot F_{\ttbar}^{MC}(i,j) \cdot P_{I/FSR}(i,j,\xi_{I/FSR}) \cdot \notag \\
& P_{Btag}(i,j,\xi_{Btag}) \cdot P_{Mistag}(i,j,\xi_{Mistag}) \cdot \notag \\
& P_{JES}(i,j,\xi_{JES}) \\
%
%
%
N_{W\bbbar}^{pred}(i,j) &= K_{W\bbbar} \cdot \sigma_{W\bbbar}^{MC} \cdot L \cdot S_{\sigma_W} \cdot F_{W\bbbar}^{MC}(i,j) \cdot \notag \\ 
& P_{Btag}(i,j,\xi_{Btag}) \cdot P_{Mistag}(i,j,\xi_{Mistag}) \cdot \notag \\ 
& P_{JES}(i,j,\xi_{JES}) \cdot P_{Q^2}(i,j,\xi_{Q^2}) \label{eqn:templateswbb} 
%
%
\end{align}
where $\sigma_\ttbar$ is the cross section and $K_{W\bbbar}$ is the relative 
normalization factor; 
$\sigma_x^{MC}$ is the cross section from 
\MC simulations; 
$L = \Lint$ is the integrated luminosity; 
$F^{MC}_{x}(i,j)$ is the \MC prediction for the fraction of events with $i$ $b$ tags and $j$ jets, 
including reconstruction and selection efficiencies; 
%
$S_{\sigma_W}$ is a factor of 1.54 obtained from the ratio of the measured \Wj cross 
section~\cite{wjetsmeasure} 
to the \alp-prediction cross section
---this is necessary due to \alp being a leading-order event generator;
and $P_{x}(i,j, \xi_x)$ are functions for each source of systematic uncertainty, $x$. 
Normalizations of the other five samples are obtained in a similar manner.
For reference and easier comparison to other measurements, 
the cross sections  
for $K = 1$ are calculated from \alp as 2744.1~pb for \Wlf, 31.9~pb for $Wc$, 
13.1~pb for $W\ccbar$, and 6.8~pb for $W\bbbar$.

\begin{figure}[htbp]
 \centering
    \resizebox{0.9\linewidth}{!}{\includegraphics{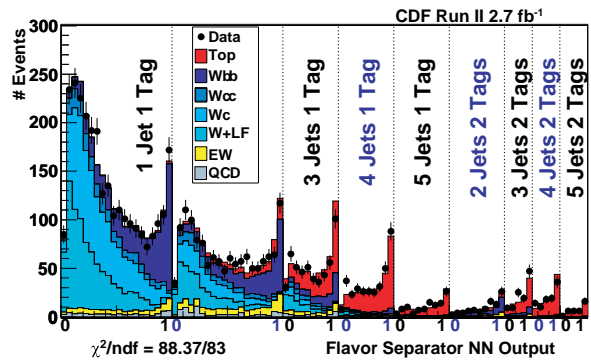}}
  \caption{(Color) The data and best fit for the flavor separator distribution for samples
  defined by the number of jets and number of tags. The legend is located in the ``2 Jets 1 Tag" bin. 
  Note that the flavor separator distribution is divided into fewer bins for 
  samples with fewer events. For samples with two or more tags, we show the average
  of the flavor separator output from the two highest-\pt tagged jets.}
  \label{fig:fitterOut}
\end{figure}

The data and best fit to the flavor separator distribution are shown 
in Fig.~\ref{fig:fitterOut}.  The
$t\bar{t}$ production cross section is found to be 
$\sigma_{\ttbar}$ = \topxsecassym~pb, and relative normalization factors are 
$K_{W\bbbar}$ = \wbbkassym, 
$K_{W\ccbar}$ = \wcckassym, 
$K_{Wc}$ = \wckassym, 
$K_{W+LF}$ = \wqqkassym, 
$K_{EW}$ = \ewkassym, and 
$K_{QCD}$ = \qcdkassym.
These results include statistical and systematic uncertainties, but do not include an uncertainty 
due to the luminosity. The top cross section we measure is consistent with theoretical
predictions~\cite{topxsectheory, topxsectheorytwo, topxsectheorythree, topxsectheoryfour}, 
and the values of the relative normalizations are consistent with what is 
expected from the theoretical uncertainty of the leading-order cross sections used by the
\MC simulation generators~\cite{alpgen, mlmmatch1}.
%

In order to evaluate the performance 
of this new method, we have compared the estimates of the systematic 
uncertainties on the top cross section with the previous method of 
background estimation~\cite{topxsecmethtwo}, though 
as applied to the same integrated luminosity. A summary of these comparisons is shown
in Table~\ref{tab:syst}. 
The total uncertainty drops from 0.84~pb to 0.73~pb, which is a 13\% improvement. However, 
the previous result developed a normalization to the $Z$ cross section to reduce the
luminosity uncertainty dramatically, and this method can be extended in the future to include 
that improvement. Therefore, upon excluding the luminosity uncertainty in order to better 
compare the methods, the uncertainty drops from 0.72~pb to 0.57~pb for a 21\% improvement.

\begin{table}
\caption{Comparison of systematic uncertainties between this result and the previous 
method of background estimation~\cite{topxsecmethtwo}. 
}
\label{tab:syst}
\begin{tabular}{|r| c |c| l}
\hline \hline
Uncertainty  & Previous Method & This Result \\
\hline
Statistical     & 0.36 pb & 0.33 pb \\
HF K-Factor     & 0.27 pb & Inc in stat \\
$Q^2$     & N/A & 0.21 pb  \\
B Tagging      & 0.39 pb  & 0.23 pb \\
Mistags & 0.17 pb & 0.08 pb \\
JES     & 0.29 pb & 0.29 pb \\
ISR/FSR      & 0.06 pb & 0.01 pb \\
Parton Showering & 0.21 pb & 0.11 pb   \\
QCD Shape & 0.06 pb & 0.01 pb \\
Flavor Separator Correction & N/A & 0.10 pb \\
Color Reconnection & N/A & 0.03 pb \\
PDF & 0.04 pb & 0.05 pb \\
Lepton ID / trigger & 0.04 pb & 0.05 pb \\
Z0 & 0.02 pb & 0.02 pb \\
\hline
Sub-Total &  0.72 pb & 0.57 pb \\
\hline
Luminosity & 0.43 pb & 0.45 pb \\
\hline
Total & 0.84 pb & 0.73 pb \\
\hline \hline
\end{tabular}
\end{table}

In summary, we measured the top pair production cross section using a novel method for 
estimating background contributions with CDF~II data corresponding to an integrated 
luminosity of 2.7~\invfb. The cross section we 
measure, \topxsecfull, is consistent with the standard model next-to-leading order
theoretical calculation~\cite{topxsectheorytwo}, and the background 
contributions are consistent with other predictions~\cite{topxsecmethtwo}. Compared 
to the previous method of background estimation, using 
$b$-tagging, this new method 
improves the precision on the top quark pair production
cross section by 21\%, excluding luminosity uncertainties.

\begin{acknowledgments}
  We thank the Fermilab staff and the technical staffs of the participating institutions for 
  their vital contributions. This work was supported by the U.S. Department of Energy and 
  National Science Foundation; the Italian Istituto Nazionale di Fisica Nucleare; the Ministry 
  of Education, Culture, Sports, Science and Technology of Japan; the Natural Sciences 
  and Engineering Research Council of Canada; the National Science Council of the 
  Republic of China; the Swiss National Science Foundation; the A.P. Sloan Foundation; 
  the Bundesministerium f\"ur Bildung und Forschung, Germany; the Korean World Class 
  University Program, the National Research Foundation of Korea; the Science and 
  Technology Facilities Council and the Royal Society, UK; the Institut National de 
  Physique Nucleaire et Physique des Particules/CNRS; the Russian Foundation for 
  Basic Research; the Ministerio de Ciencia e Innovaci\'{o}n, and Programa 
  Consolider-Ingenio 2010, Spain; the Slovak R\&D Agency; and the Academy of Finland.
\end{acknowledgments}

\bibliography{top_cdf}

\begin{thebibliography}{10}
\bibitem[9]{coords}
We use a cylindrical coordinate system with its origin in the center of the
  detector, where $\theta$ and $\phi$ are the polar and azimuthal angles,
  respectively, and pseudorapidity is $\eta = -\ln \tan(\theta/2)$. The missing
  \et ($\vec{\met}$) is defined by $\vec{\met} = - \sum_i \et^i n^i, i =$
  calorimeter tower number, where $n^i$ is a unit vector perpendicular to the
  beam axis and pointing at the $i^{th}$ calorimeter tower. $\vec{\met}$ is
  corrected for high- energy muons and also jet energy corrections. We define
  $\met = |\vec{\met}|$. The transverse momentum, \pt, is defined to be
  $p\sin\theta$.
\end{thebibliography}

\end{document}